\documentclass[aps,prl,twocolumn]{revtex4}
\usepackage{graphics}

\makeatletter

\providecommand{\LyX}{L\kern-.1667em\lower.25em\hbox{Y}\kern-.125emX\@}

\usepackage{bm}

\input epsf

\def\be{\begin{equation}}
\def\ee{\end{equation}}
\def\ba{\begin{eqnarray}}
\def\ea{\end{eqnarray}}

\makeatother
\begin{document}

\title{Vortex liquid crystals in anisotropic type II superconductors}

\author{E.~W.~Carlson\( ^{1,2} \), A.~H.~Castro Neto\( ^{1} \), and
D.~K.~Campbell\( ^{1,2} \)}

\address{(1) Dept. of Physics (2) Dept. of Electrical and Computer Engineering
\\ Boston University, Boston, MA 02215}

\date{\today}

\begin{abstract}
In a type II superconductor in a moderate magnetic field, the superconductor
to normal state transition may be described as a phase transition
in which the vortex lattice melts into a liquid. In a biaxial superconductor,
or even a uniaxial superconductor with magnetic field oriented perpendicular
to the symmetry axis, the vortices acquire elongated cross sections
and interactions. Systems of anisotropic, interacting constituents
generally exhibit liquid crystalline phases. We examine the possibility
of a two step melting in homogeneous type II superconductors with
anisotropic superfluid stiffness from a vortex lattice into first
a vortex smectic and then a vortex nematic at high temperature and
magnetic field. We find that fluctuations of the ordered phase favor
an instability to an intermediate smectic-A in the absence of intrinsic
pinning.
\end{abstract}

\pacs{74.60.Ec, 74.60.Ge, 61.30.-v}

\maketitle
Recently, there has been much interest generated concerning high temperature
superconductors in a magnetic field. Various experiments have studied
the interplay of superconductivity with coexistent magnetic orders
in the presence of an external field. Experiments using both neutron
scattering\cite{lakenature,lakescience} and STM\cite{hoffman} show
that there is significant local electronic inhomogeneity. These and
other experiments lend credence to the idea that there may be electronic
liquid crystalline phases in strongly correlated systems, leading
to anisotropy even within a CuO\( _{2} \) plane. 

The cuprate superconductors are also ideal laboratories for studying
vortex physics, due to the large values of \( \kappa \equiv \lambda _{ab}/\xi _{ab} \)
(where \( \lambda _{ab} \) and \( \xi _{ab} \) are the London penetration
depth within a plane, and the coherence length within a plane, respectively)
and small critical depinning current.\cite{blatter} In this letter,
we consider the effects of an anisotropic superfluid stiffness on
the vortex phases in superconductors in the continuum limit. We account
for this anisotropy by allowing different effective masses in the
three crystalline directions, which we will call \( m_{a}, \) \( m_{b}, \)
and \( m_{c}. \)

\begin{figure}
{\centering \resizebox*{!}{6.5cm}{\includegraphics{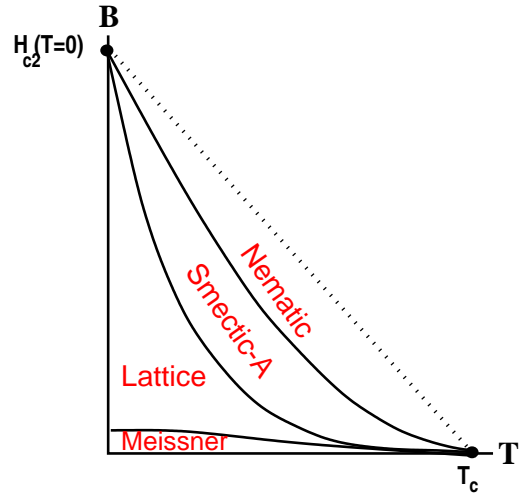}} \par}

\caption{Schematic phase diagram of vortex matter with anisotropic interactions.
Solid lines represent phase transitions, and the dotted line represents
the crossover at \protect\( H_{c2}.\protect \) There may also be
melted or partially melted phases near \protect\( H_{c1},\protect \)
between the regions marked Meissner and Lattice. \label{fig:phase}}
\end{figure}

In a biaxial superconductor (with different effective mass in each
crystalline direction), or a uniaxial superconductor (the effective
mass differs in one direction only) with magnetic field oriented perpendicular
to the symmetry axis, vortices acquire elliptical cross sections,
whether measured by the shape of the core, or by the profile of the
screening currents or magnetic field which penetrates beyond the core.
Systems of anisotropic interacting constituents generically lead to
liquid crystalline phases. In such a system, we expect the melting
to proceed from the body centered rectangular lattice, to a smectic,
to a nematic, as in Fig. \ref{fig:phase}. (In this case, the high
temperature phase is trivially nematic due to the explicitly broken
rotational symmetry introduced by the mass anisotropy.) 

\begin{figure}
{\centering \resizebox*{0.75\columnwidth}{!}{\includegraphics{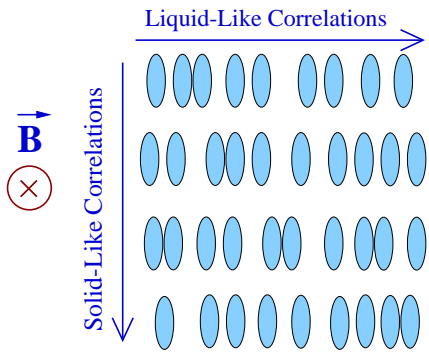}} \par}

\caption{{\it Smectic-A.} Cross sections of vortices in an anisotropic superconductor
are represented schematically above as filled ellipses, for a magnetic
field oriented perpendicular to the page. There may be an intermediate
melting from the elongated lattice first to a smectic-\protect\( A,\protect \)
shown above, and then to a nematic phase at high temperature.\label{fig:xsec}}
\end{figure}

Liquid crystals lie somewhere between the full translational and rotational
symmetry of a liquid, and that of a 3D crystal, which has broken rotational
symmetry, and retains only discrete translational symmetry in the
three directions of the crystal axes. 
 In a superconductor, the application of an external magnetic field
to produce vortices explicitly breaks rotational symmetry. We choose
axes such that \( B || \widehat{z.} \) In the vortex system,
smectic phases correspond to liquid-like correlations (and unbroken
translational symmetry) in one direction in the \( xy \) plane, and
simultaneous solid-like correlations (and only discrete translational
symetry) in the other direction in the \( xy \) plane. Smectics may
be further classified by which direction the {}``elongated molecule''
is pointing on average with respect to the orientation of the liquid-like
layers. Call \( \theta  \) the angle that the long axis of the molecule
makes with respect to the normal of the liquid-like layers. For \( \theta =0, \)
the phase is smectic-\( A \), illustrated in Fig. \ref{fig:xsec}.
For all other values of \( \theta \,  \) the phase is a smectic-\( C \).
A nematic phase is characterized by unbroken translational symmetry,
with broken orientational symmetry. Correlations in this phase are
liquid-like in all directions, but the constituent molecules have
a preferred orientation.

When the magnetic field is oriented parallel to the planes\cite{efetov,balentsprb,balentsprl}
in a layered superconductor, the explicit translational symmetry breaking
of the planes may cause the vortex lattice to melt first along the
direction of the planes, leading to a smectic-\( C \) with \( \theta =\frac{\pi }{2}. \)
 Smectic phases have also been predicted in the presence of a driving
current\cite{radzihovsky}, as well as chain states which may arise
when the field is tilted sufficiently away from a crystalline axis\cite{chain1,chain2}.
Here we explore the possibility of liquid crystalline phases due to
explicit rotational symmetry breaking (mass anisotropy), with no explicit
translational symmetry breaking in the problem ({\it i.e.} no intrinsic
pinning). We find that thermal fluctuations of such an anisotropic
vortex lattice favor an instability to a vortex smectic-\( A \).
Such a phase, if it exists, should be detectable by several experimental
probes, including Bitter decoration experiments, neutron scattering,
\( \mu  \)SR, and resistivity measurements.

We use continuum elasticity theory to describe the thermal fluctuations
of an ordered vortex lattice in three dimensions. We choose axes such
that the external magnetic field \( B||\widehat{z}. \) The displacement
of a vortex from its equilibrium position is denoted by the vector
\( \mathbf{u}=\left( u_{x},u_{y}\right) , \) which is a function
of the position \( z \) along a given vortex. To second order in
\( \mathbf{u}, \) the free energy is \begin{equation}
\label{free_energy}
F=\frac{1}{2}\int \frac{d^{3}k}{(2\pi )^{3}}\mathbf{u}\cdot \mathbf{C}\cdot \mathbf{u},
\end{equation}
 where the matrix \( \mathbf{C} \) contains the elastic constants:\cite{blatter}

\begin{equation}
\label{Cxy}
\mathbf{C}=\left( \begin{array}{cc}
c_{11}k_{y}^{2}+c_{66}^{h}k_{x}^{2}+c_{44}^{h}k_{z}^{2} & c_{11}k_{x}k_{y}\\
c_{11}k_{x}k_{y} & c_{11}k_{x}^{2}+c_{66}^{e}k_{y}^{2}+c_{44}^{e}k_{z}^{2}
\end{array}\right) .
\end{equation}
 Here we orient the magnetic field perpendicular to the axis of symmetry:
\( \overrightarrow{B}\perp \widehat{c}||\widehat{z}. \) We have also
assumed a uniaxial superconductor, \( m_{a}=m_{b}\equiv m_{ab}\neq m_{c} \)
; the elastic constants are not yet known for the fully anisotropic,
biaxial case. Nonetheless, this uniaxial geometry captures the physics
we are interested in, namely anisotropic interactions. We use elastic
constants derived from Ginzburg-Landau theory.\cite{sudboc4,squash,blatter}
The bulk modulus, \( c_{11}(\mathbf{k}) \), describes the compressibility
of the lattice. The hard tilt modulus, \( c_{44}^{h}(\mathbf{k}) \),
corresponds to tilts along the symmetry axis \( \widehat{c}, \) and
the (smaller) easy tilt modulus, \( c_{44}^{e}(\mathbf{k}) \), corresponds
to tilts perpendicular to \( \widehat{c}. \) Similarly, it is easier
to shear vortices (\( c_{66}^{e} \)) along the major axis of the
cross sectional ellipse, rather than perpendicular to it (\( c_{66}^{h} \)).
Note that since the magnetic field is oriented along a crystal symmetry
axis, there is no mixing between the bulk, tilt, and shear moduli.
The bulk and tilt moduli are highly momentum dependent. In fact, the
two soften significantly at the Brillouin zone edge, so that their
momentum dependence is important in the physics of melting. The shear
moduli are approximately independent of the wavelength of the distortion,
and we neglect their weak momentum dependence. 

We use an extension of the Lindemann criterion to the case of anisotropy\cite{balentsprb}
and allow for the possibility that the lattice may melt in one direction
before the other. In this case, fluctuations in the \( x \) direction
compete with the lattice spacing in the \( x \) direction, and fluctuations
in the \( y \) direction with the lattice spacing in the \( y \)
direction:\begin{eqnarray}
<u_{x}^{2}>= & \frac{1}{2}c^{2}a^{2}\gamma ^{2} & \nonumber \\
<u_{y}^{2}>= & \frac{1}{2}c^{2}a^{2}/\gamma ^{2},\label{lindemann} 
\end{eqnarray}
 where \( \gamma ^{4}\equiv \frac{m_{ab}}{m_{c}}, \) \( a=\sqrt{\frac{2\Phi _{o}}{3^{1/2}B}} \)
is the lattice spacing for the triangular lattice of the isotropic
case at the same magnetic field strength \( B \), and \( \Phi _{o} \)
is the quantum of flux. We look for this criterion to be significantly
violated in one direction before the other. The factor of \( 1/2 \)
allows the Lindemann parameter c to recover the usual definition in
the isotropic case.

For short wavelengths, the physics of an anisotropic superconductor
can be mapped onto an isotropic superconductor by a scaling procedure
introduced by Blatter {\it et al.} \cite{scaling} Were this true
at all wavelengths, there would be no reason to expect anisotropic
melting to occur. However, scaling breaks down for the long wavelength
bulk and tilt modes of the system, which are isotropic and have a
vanishing energy cost. The result is that the spatial profile of the
fluctuations of the vortices is less eccentric than the equilibrium
lattice, suggesting an instability to partially melted (liquid crystalline)
phases. 

Using the scaled momenta, \( \mathbf{q}=(\gamma k_{x}/\Lambda ,k_{y}/\gamma \Lambda ,k_{z}/\Lambda ), \)
the average fluctuations may be written as: \begin{eqnarray}
<u^{2}_{x}> & = & \frac{\Lambda }{B^{2}}\frac{k_{B}T}{(2\pi )^{3}}\int d\mathbf{qC}^{-1}_{xx}(\mathbf{q})\\
<u_{y}^{2}> & = & \frac{\Lambda }{B^{2}}\frac{k_{B}T}{(2\pi )^{3}}\int d\mathbf{qC}^{-1}_{yy}(\mathbf{q})
\end{eqnarray}
where the matrix \( \mathbf{C}\left( \mathbf{q}\right)  \) and the
elastic constants therein are functions of \( \mathbf{q}, \) and
the cutoff \( \Lambda = \) \( \sqrt{\frac{4\pi B}{\Phi _{o}}}\propto \frac{1}{a} \)
is set by the vortex lattice spacing. The integrals are functions
only of \( \kappa , \) \( \gamma , \) and \( b\equiv \frac{B}{H_{c2}(T)}. \)
We compute the integrals numerically to obtain the melting curves. 

\begin{figure}
{\centering \resizebox*{1\columnwidth}{!}{\rotatebox{270}{\includegraphics{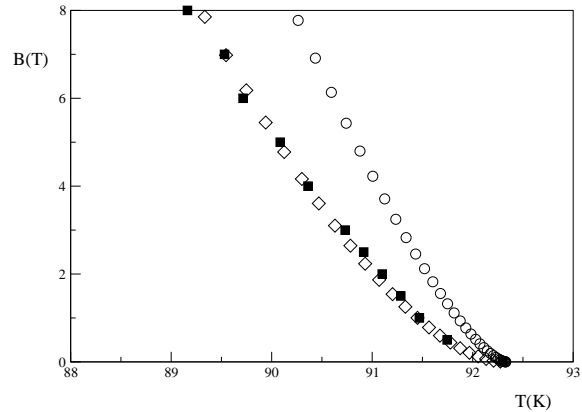}}} \par}

\caption{Solid squares are vortex lattice melting data on optimally doped
YBCO with \protect\( B\perp c\protect \)\cite{kwok}. Open symbols
are the results of the numerical integration of Eqn. \ref{lindemann},
taking into account the pinning of the planes. Circles refer to melting
in the {}``short'' direction, and diamonds to melting in the {}``long''
direction. We have taken the following parameters, appropriate for
optimally doped YBCO with \protect\( B\perp c\protect \) : \protect\( T_{c}=92.3K,\protect \)
\protect\( \gamma ^{-4}=\frac{m_{c}}{m_{ab}}=59,\protect \) \protect\( \kappa =\frac{\lambda _{ab}}{\xi _{ab}}=55,\protect \)
and \protect\( H_{c2}^{c}=842T.\protect \) The figure is plotted
for \protect\( c=.19\protect \) . \label{kwokdata}}
\end{figure}

\begin{figure}
{\centering \resizebox*{0.9\columnwidth}{5.5cm}{\rotatebox{270}{\includegraphics{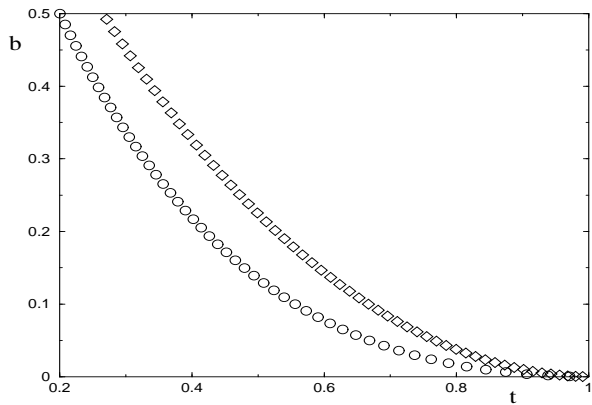}}} \par}

\caption{Results of the numerical integration of Eqn. \ref{lindemann}, in
the absence of pinning. The circles refer to melting in the {}``short''
direction, and the diamonds to melting in the {}``long'' direction.
We have taken the following parameters: \protect\( \gamma ^{-4}=\frac{m_{c}}{m_{ab}}=10,\protect \)
\protect\( \kappa =\frac{\lambda _{ab}}{\xi _{ab}}=100,\protect \)
and \protect\( H_{c2}^{c}(T=0)=100T.\protect \) The figure is plotted
for \protect\( c=.2\protect \), as a function of \protect\( b=\frac{B}{H_{c2}(T)}\protect \)
and \protect\( t=\frac{T}{T_{c}}.\protect \) \label{meltA}}
\end{figure}

We first compare to data on optimally doped YBCO, with \( B\perp c. \)
It is believed that in this geometry, the intrinsic pinning of the
planes leads to a partially melted phase which is a smectic-\( C \),
in which vortices have melted along the planes. In the case of a lattice
which is commensurate with the planes, the explicit symmetry breaking
of the planes adds a momentum-independent pinning term to the matrix
element \( \mathbf{C}_{yy} \) \cite{ivlev1990} and tends to encourage
melting along the planes, into a smectic-\( C \). In this sense,
pinning competes with the aforementioned tendency of the anisotropy
to encourage a smectic-\( A \). We find for this material that the
Lindemann criterion is violated in one direction well before the other,
as shown in Fig. \ref{kwokdata}, and the instability favors a smectic-\( C. \)
For an incommensurate lattice, pinning is irrelevant, in which case
the melting may proceed as in Fig. \ref{meltA} as discussed below.
For optimally doped YBCO with \( B\perp c \), the lattice remains
commensurate up to at least 120 Tesla, but in other types of superconductors
it may be possible to reach incommensurability at lower fields.\cite{thompson}
Note that our approach is only capable of calculating the first melting
curve, from solid to smectic. To calculate the melting curve for smectic
to nematic, it is necessary to first derive elastic constants for
the smectic phase, which is beyond the scope of the present paper. 

In Fig. \ref{meltA}, we plot the results {\it in the absence of intrinsic pinning.} The
instability now favors a smectic-\( A, \) in which the long direction
of each constituent {}``molecule'' (in this case, the major axis
of each cross sectional ellipse of a vortex) is oriented perpendicular
to the melted layers, as in Fig. \ref{fig:xsec}. Although this means
that the vortices have melted first along the direction of harder
shear, \( c^{h}_{66}, \) this is the most common smectic geometry
observed for oblong molecules. We show the schematic phase diagram
in Fig. \ref{fig:phase}. The smectic region may be pinched off by
first order transitions from lattice directly to nematic near \( H_{c2}(T=0) \)
and near \( T_{c}. \) 

The problem of vortices in a 3D superconductor may be mapped to that
of 2D bosons at zero temperature, by approximating the vortex interactions
as {}``local'' in the coordinate \( \hat{z} \), where the path
of the vortex represents a bosonic world line. The theory of anisotropic
2D melting in the presence of explicitly broken rotational symmetry
predicts a region of quasi-long-range ordered (QLRO) smectic-\( A \)
at finite temperature.\cite{ostlundhalperin,ostlundanis} At zero
temperature (to which the current case maps), a long range ordered
(LRO) smectic-\( A \) is possible. The intermediate smectic-\( A \)
has also been seen in recent numerical simulations of a two dimensional
vortex system\cite{lanlsim}. 

For the present case, the soft rotational modes usually responsible
for preventing translational LRO in the smectic are absent because
the mass tensor introduces explicit rotational symmetry breaking.
It costs energy to rotate the vortex smectic, and the system exhibits
gradient elasticity. It follows that a 3D smectic with explicitly
broken rotational symmetry can have translational LRO. An interesting
consequence of this is that the rigidity of the vortex smectic preserves
superconductivity between the liquid-like layers, so that the transition
from vortex lattice to vortex smectic is also a transition from 3D
superconductivity to 2D superconductivity.

Although we have presented results for a (homogeneous) uniaxial superconductor,
we also expect the results to apply for a biaxial superconductor,
with three different entries in the mass tensor. The cuprates certainly
exhibit anisotropy between the \( c \) direction and the planar directions,
but they often also exhibit anisotropy within the \( ab \)-plane.
In particular, our assumptions of mass anisotropy with no explicit
translational symmetry breaking in the electronic degrees of freedom\cite{crystal}
are in principle satisfied for the geometry \( B||c \) in the cuprates
in the presence of an electron nematic phase\cite{kivelfrad} within
the planes. Our assumptions may also be satisfied in stripe ordered
phases, provided the mutual pinning is not too strong and thermal
depinning of vortices from stripes occurs at a lower temperature than
that at which the vortex lattice melts.  

The smectic-\( A \) has clearest implications for experimental probes
that are capable of measuring the structure function of the vortex
order, such as Bitter decoration (which is surface sensitive) or neutron
scattering (which is a bulk probe). In melting from lattice to smectic,
the diffraction pattern loses most Bragg peaks, retaining at most
a line of Bragg peaks along the direction of modulated density, in
this case a modulation of average magnetic field density. More commonly,
smectics exhibit the central Bragg peak along with a pair of peaks
associated with the first harmonic of the density modulation.

There are also distinctive implications of the double melting for
\( \mu SR \). Muon spin rotation detects the distribution of local
magnetic fields. In the nematic phase, the time-averaged magnetic
field density is uniform. In partialy freezing from the nematic to
the smectic, \( \mu SR \) would exhibit a new inhomogeneity in the
magnetic field in the smectic state. Upon freezing further into the
vortex lattice, the \( \mu SR \) signal would reveal another transition
to further magnetic inhomogeneity. The changes in the \( \mu SR \)
signal are expected to coincide with the onset of highly anisotropic
resistivity in going to the smectic, and with the onset of 3D superconductivity
upon entering the lattice phase. 

Resistivity measurements are also sensitive to smectic order. When
the Lorentz force is along a liquid-like direction, vortices move
easily and the resistivity is large. When the Lorentz force is along
the solid-like direction, the rigidity of the smectic resists vortex
motion, and the resistivity vanishes (although the movement of defects
may provide some small amount of dissipation).\cite{balentsprl} If
the current is along the magnetic field direction, the Lorentz force
and dissipation are negligible. The vortex smectic-\( A \) retains
2D superconductivity between the liquid-like layers of vortices, with
the resistivity \( \rho _{||} \) vanishing parallel to the layers,
but \( \rho _{\perp }\neq 0 \) for currents applied perpendicular
to the smectic layers. 

In conclusion, we have studied the problem of vortex lattice melting
in anisotropic superconductors in the continuum limit. The introduction
of anistropy in the mass tensor leads to elongation of vortex cross
sections and interactions. We have demonstrated that interacting elongated
vortices can form liquid crystalline phases. Using elasticity theory,
with momentum-dependent elastic constants derived from Ginzburg-Landau
theory, we have calculated the thermal fluctuations of the vortex
lattice. Comparing these results to an anisotropic Lindemann criterion,
we argue that there is an instability to an intermediate smectic phase.
In the absence of intrinsic pinning, we find an instability favoring
a smectic-\( A, \) wherein the lattice has melted along the direction
of the shorter lattice constant. 

It is a pleasure to acknowledge fruitful discussions with R. A. Pelcovits,
V. G. Kogan, T. C. Lubensky, L. Balents, L. Radzihovsky, G. Blatter,
H. A. Mook, C. Reichhardt, C. Olson, and D. Nelson. This work was
supported in part by NSF grant DMR-97-12765 (EWC and DKC) and the
Office of the Provost at Boston University (EWC). 

\bibliographystyle{notitles2}
\bibliography{smectic}

\end{document}